# Preparation of Sol-Gel Random Micro Lens Array

Fanru Kong[1], Chuanzhu Cheng[1], Yuqing Liu*

*Center for Advanced Optoelectronic Functional Materials Research, and Key Laboratory for UV-Emitting Materials and Technology of Ministry of Education, National Demonstration Center for Experimental Physics Education, School of Physics, Northeast Normal University, 5268 Renmin Street, Changchun 130024, Jilin, China*

Abstract: The structure of random micro lens array (rMLA) breaks the periodicity of micro lens array (MLA), suppressing coherence in the homogenization process, thereby achieving better spot homogenization effects. Sol-gel rMLA exhibits strong adaptability and high laser tolerance, making it valuable for laser beam control applications. However, the cracking tendency during the drying process of sol-gel is a challenge. This paper successfully prepares sol-gel random micro lens arrays through nanoimprint lithography, thoroughly analyzing the cracking mechanism and resolving the cracking issue during the drying process of sol-gel. The manufactured sol-gel random micro lenses exhibit good surface profile accuracy, uniformity, and excellent light source shaping effects. The energy utilization efficiency of various types of rMLA is approximately 90%, with rectangular and hexagonal rMLAs achieving uniformity of light spots of over 80%.
Keywords: Sol-gel; Random micro lens array; Digital microlens devices

## 1.Introduction

In recent years, the widespread application of lasers in fields such as laser illumination, laser detection, and communication has attracted considerable attention. However, laser light typically exhibits a Gaussian distribution with high central intensity and weak edges, which is not suitable for use in lithography, laser processing, and high-performance lighting. To address this issue, various methods have been proposed, including non-spherical lens groups, freeform lenses, diffractive optical elements (DOE), and micro lens arrays (MLA). However, each of these methods has its limitations. Among them, micro lens arrays (MLA) are considered one of the most commonly used homogenizers due to their high energy utilization efficiency, flexibility, and uniformity. However, due to the periodic structure of MLA, interference occurs between small beams, resulting in interference fringes appearing in the homogenized light spot, thus reducing uniformity. To solve this problem, several methods have been proposed, such as randomly arranging the sub-lenses of the lens array to eliminate interference effects on the homogenized light spot. While these methods have been theoretically proven feasible, current processing techniques are not mature enough to reliably mass-produce this structure.

Sol-gel is a material formed by the chemical reaction of precursors under hydrolysis and condensation in the liquid phase. By adjusting parameters such as precursor concentration and acidity, the morphology and microstructure of the material can be precisely controlled, resulting in more stable material properties. Complex-shaped devices can be fabricated through precision processing methods such as compression molding. Common materials for the preparation of optical elements currently include glass, resin, and sol-gel. Since glass materials require firing, the long production cycle and high manufacturing costs make glass material optical element preparation less favorable. In contrast to traditional glass manufacturing techniques, sol-gel does not require firing, resulting in lower cost, simpler operation, shorter production cycles, and easier mass production. Moreover, due to the shrinkage of glass materials during firing, it is difficult to accurately control this shrinkage rate, resulting in deviations between glass material optical elements and design values. Compared to optical devices made from resin, optical devices prepared from sol-gel have higher mechanical strength and stability. The common melting points of resin materials are only 80°C-120°C, and they are prone to decomposition at high temperatures. It is precisely because of this property that optical devices made from resin are not suitable for high-power lasers.

The sol-gel material configured in this paper is simple to prepare and has a high glass transition temperature. Through testing, the glass transition temperature of the sol-gel material is determined to be 274°C. When subjected to 40W high-power laser irradiation, there is no significant change in the shape and morphology of the random micro lens's light spot, indicating that sol-gel has good temperature resistance and is suitable for high-power lasers. Sol-gel random micro lens arrays manufactured through specific processes exhibit good surface profile uniformity and accuracy. Combined with the hydrolysis principle of sol-gel, sol-gel capable of photopolymerization is prepared, and based on this, square and hexagonal random micro lens arrays are fabricated through specific processing techniques. After testing, the random micro lens arrays exhibit excellent light source shaping effects, with an energy utilization rate of up to 90% and light spot uniformity of over 80%.

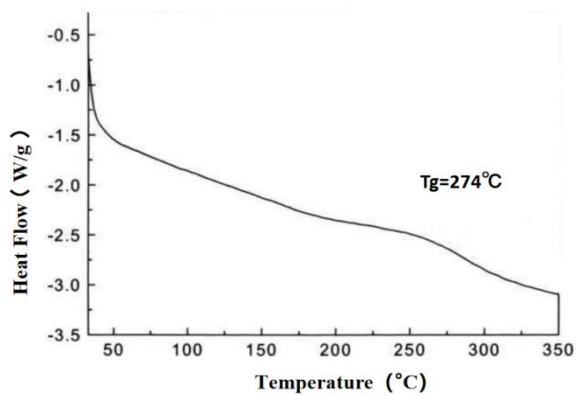

Figure 1: Relationship between sol-gel thermal flux and temperature.

## 2 .Preparation of sol-gel

The sol-gel technique is a new process for obtaining oxides or other compounds from mixtures of metal organic compounds, metal inorganic compounds, or both, through a process of hydrolysis and condensation, gradual gelation, and corresponding post-treatment. Using liquid chemical reagents as raw materials, these materials are uniformly mixed in the liquid phase to form a series of chemical reactions such as hydrolysis and condensation (polymerization) in the solution to form a stable transparent sol solution. The sol undergoes slow polymerization between gel particles through aging to form a gel, which is then dried to remove the solvent, resulting in a dry gel or aerogel with a porous spatial structure, which is then prepared by sintering and solidification.

The sol-gel consists of two main components: a $SiO_2$ network composed of MAPTMS hydrolysis and condensation and a $TiO_2$ network composed of titanium isopropoxide hydrolysis. Among them, titanium isopropoxide acts to adjust the refractive index of the liquid and improve the resolution of MAPTMS photopolymerization. In this experiment, Solution I is formed by hydrolyzing methyl methacrylate propyltrimethoxysilane (MAPTMS) in isopropanol to form the silica network. 3 ml of MAPTMS was added to a beaker, followed by the addition of 1.2 ml of isopropanol and 0.2 ml of water. After stirring the mixture for 1 hour, a clear solution was obtained. Solution II is a titanium dioxide network, obtained by mixing 1.2 ml of titanium isopropoxide and 1.8 ml of acetone and stirring. After stirring Solution II for 1 hour, a yellow solution was obtained. Solutions I and II were mixed and stirred at room temperature for a certain period of time, and an appropriate amount of hydrochloric acid was added. 3% by weight of photoinitiator 819 and 1% by weight of photoinitiator 1-hydroxycyclohexyl phenyl ketone were added to the mixture of Solution I and Solution II. This makes the material UV-sensitive. The corresponding preparation process is shown in diagram (e). The decomposition and coupling process of the photoinitiator are shown in (b). Under UV irradiation, the photoinitiator 1-hydroxycyclohexyl phenyl ketone (C13H16O2) can decompose into two free radicals that can polymerize methyl methacrylate groups or produce new molecules through coupling of two benzoyl radicals. When the two free radicals of the photoinitiator polymerize with MAPTMS, new molecules are formed, as shown in diagram ©. Photoinitiators are added to the sol-gel system, and when exposed to UV light, the photoinitiator decomposes to generate free radicals. These free radicals then combine with the C=C double bonds contained in the methyl methacrylate groups of MAPTMS, triggering the polymerization reaction. In the initial stage of illumination, these free radicals combine with the C=C double bonds, initiating the polymerization process, but their rate of generation is significantly lower than their rate of consumption. Subsequently, as monomers need to be transferred, the rate of combination of free radicals and C=C double bonds decreases, and the rate of consumption of free radicals also decreases. After a very short period of time, the rates of generation and consumption of free radicals will gradually balance, indicating that the reaction enters a steady state, where the rates of generation and consumption of free radicals are equal. In other words, the concentration of C=C double bonds is an important factor in determining whether sol-gel photopolymerization can occur. Diagram (d) shows the process of sol-gel hydrolysis and condensation.

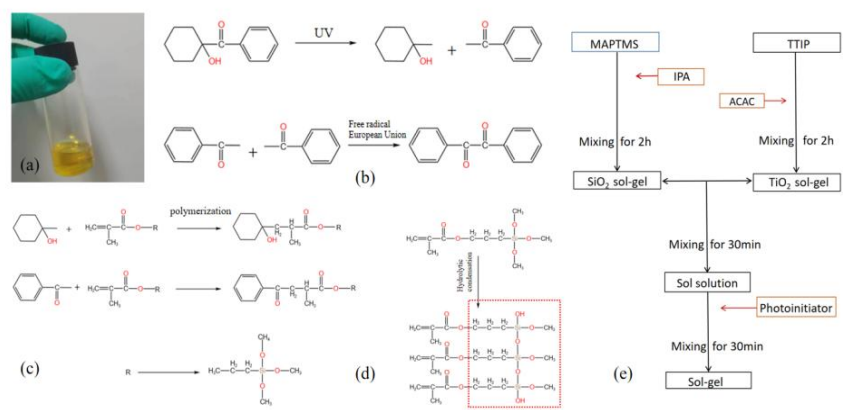

Figure 2 (a) Configured sol-gel; (b) Photoinitiator decomposition and coupling;(c) Polymerization process; (d) Hydrolysis-condensation process; (e) Configuration process.

3.Production of random microlenses via sol-gel

3.1Analysis of cracking mechanism in sol-gel

Cracking is a phenomenon that occurs in materials during the preparation process. Sol-gel is a phase transition process from sol to gel, where sol is a colloidal system formed by dispersing solid particles in a liquid medium, and gel is a solid substance formed by gradually aggregating these solid particles. Cracking of sol-gel is usually caused by internal stress during drying, leading to the generation of cracks. Figure 3(b) shows the cracking phenomenon of sol-gel. The reasons for cracking of sol-gel are as follows: during drying, solvent evaporation enhances the interaction between particles inside the sol-gel, gradually forming a dense network. This gel contraction generates internal stress, which, if unevenly affected by temperature changes, leads to cracking. Non-uniform sol-gel structures or large gradients in water content can cause uneven volume shrinkage and cracking. Proper drying conditions are crucial to reduce sol-gel cracking. Gradually reducing drying temperature and humidity can slowly and uniformly remove solvent, preventing the accumulation of internal stress. Additionally, precursor concentration affects cracking. Proper heating and stirring of sol-gel materials before processing can increase concentration, and thinner coating during processing can alleviate cracking during drying. Finally, the drying process, which allows organic compounds to evaporate, is essential. Here, we adopt a relatively mild room temperature drying for 24 hours for random microlens array. Through this process, the cracking problem is solved, as shown in Figure 3(d). After drying, the sol-gel random microlens array will not crack.

**3.2 Preparation of random microlens arrays**

For microlens arrays, the first consideration is the shape that can be densely packed in the entire space. For any regular N-sided polygon, it satisfies the condition of dense packing if the internal angle can be evenly divided, then the following mathematical relationship holds

$$\frac{360}{\frac{N-2}{N} \times 180} \in Z$$

Solving, that is, triangles, quadrilaterals, and hexagons can be densely arranged in space. Then the boundary shape of the microlens array remains consistent with the boundary shape of its basic unit, and the method of dense arrangement is periodic, ensuring that the basic unit can only be densely packed through translation. Finally, combining with the etching process, for the basic principle of etching itself, the boundary is approximately determined by the perpendicular bisector of the line connecting two adjacent mask holes. However, the characteristics of parallelograms make them not satisfy this etching principle and cannot be used as the basic unit for etching.

Random positioning mainly involves adding a certain random variable to a regular position distribution. The design of the mask starts from a regular situation, where the position occupied by each mask hole is regular, and their sizes are also consistent. Based on this regularity, the etching method with random distribution of etching positions is adopted. That is, each mask hole is randomly moved around from its original regular position according to the set random degree, while the size of the mask hole remains unchanged, resulting in an array with regular apertures but random positions.

0.1 ml of photosensitive sol-gel is dropped onto a PDMS template and spin-coated at 600 rpm for 10 seconds. The template is then set aside for later use. 0.2 ml of sol-gel is dropped onto a square glass slide and spin-coated at 600 rpm. The PDMS template with spin-coated photosensitive sol-gel is placed on the glass slide with spin-coated sol-gel. Figure 3(e) shows the preparation process of the random microlens array, with the PDMS template on top and the square glass slide below. It is left at room temperature for 1 minute, during which capillary action causes the sol-gel to automatically fill the remaining areas. After 1 minute, the DMD light source switch is turned on, and it is illuminated for 18 seconds before being left at room temperature. After half an hour of standing, carefully remove the PDMS mold.

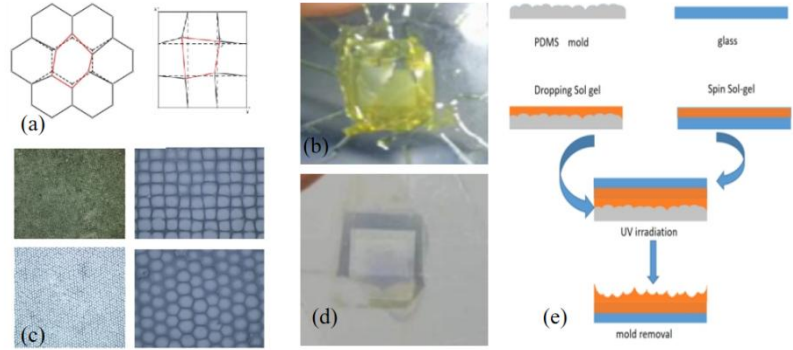

Figure 3 (a) Schematic diagram of random microlens array; (b) Cracking phenomenon of random microlens array; (c)View of prepared random microlens array under optical microscope; (d) Random microlens array without cracks; (e) Process flowchart of random microlens array fabrication.

### 3.3 Characterization and testing of random microlens arrays

Before laser beam homogenization, the three-dimensional distribution and cross-sectional intensity distribution are illustrated for the laser spot captured by the CCD. From the graph, it can be observed that the intensity curve of the laser beam follows a typical Gaussian distribution. The three-dimensional and cross-sectional intensity distributions after uniformization through a rectangular random microlens array are respectively shown. After the uniformization process, the intensity distribution of the laser beam is widened into a flat-top distribution, as shown in the following figure of the homogenized spot captured by the CCD. Similarly, the three-dimensional and cross-sectional intensity distributions after uniformization through a hexagonal random microlens array are respectively shown. After uniformization through the hexagonal random microlens array, the intensity distribution of the laser beam is widened into a flat-top distribution. The inset shows the homogenized spot captured by the CCD.

Subsequently, uniformity and energy utilization tests were conducted on the prepared rectangular and hexagonal homogenizing elements. The uniformity of the laser spot is calculated by the formula.

$$RMS = \sqrt{\sum_{j}^{N}(I_j - \overline{I})/N}$$

、

Where the uniformity of the rectangular and hexagonal spots is approximately 80% and 85% respectively. The input power of the beam measured by the power meter ($P_{in}$) is approximately, while the output power ($P_{out}$) after passing through the hexagonal and rectangular elements is respectively. According to the formula, the energy utilization rates are approximately 89% and 91% respectively.

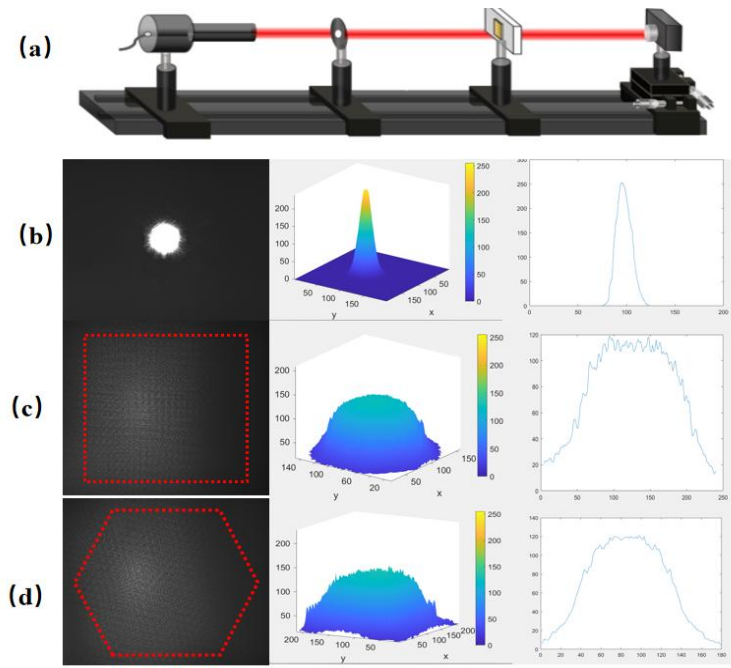

Figure 4 (a) Schematic diagram of the testing optical path; (b) Three-dimensional and cross-sectional intensity distribution of the laser beam before light source homogenization; Three-dimensional and cross-sectional light intensity distribution after homogenization by rectangular random microlens array (rMLA); (d) Three-dimensional and cross-sectional light intensity distribution after homogenization by hexagonal rMLA.

**4.summarize**

   This paper presents the preparation of rectangular and regular hexagonal multiple-type random microlens arrays based on the sol-gel method. Furthermore, a method suitable for the preparation of lens arrays is explored to address the cracking of sol-gel. By employing certain additional processes, the cracking issue of sol-gel is effectively solved. Compared to traditional glass materials, this method offers advantages such as shorter production cycles and lower costs, meeting the requirements of optical components. Rectangular and hexagonal arrays exhibit excellent homogenization effects on laser beams, with energy utilization rates exceeding 90%, and the uniformity of homogenized spots is approximately 82% and 85%, respectively